\documentclass{emulateapj}

\usepackage{graphicx}
\usepackage{float} 
\usepackage{amsmath}
\usepackage{epsfig,floatflt}

\begin{document}

\title{Spatial variations in the spectral index of polarized
  synchrotron emission \\ in the 9 yr \emph{WMAP} sky maps}

\author{U. Fuskeland\altaffilmark{1}, I. K. Wehus\altaffilmark{2,3}, H. K. Eriksen\altaffilmark{1}, and S. K. N{\ae}ss\altaffilmark{1,3}}

\email{unnif@astro.uio.no}
\email{i.k.wehus@fys.uio.no}
\email{h.k.k.eriksen@astro.uio.no}
\email{s.k.nass@astro.uio.no}

\altaffiltext{1}{Institute of Theoretical Astrophysics, University of
  Oslo, P.O.\ Box 1029 Blindern, NO-0315 Oslo, Norway}

\altaffiltext{2}{Jet Propulsion Laboratory, California Institute of Technology, Pasadena, CA 91109, USA}

\altaffiltext{3}{Astrophysics, University of Oxford, DWB, Keble Road,
  Oxford OX1 3RH, UK}


\begin{abstract}
We estimate the spectral index, $\beta$, of polarized synchrotron 
emission as observed in the 9 yr \emph{WMAP} sky maps using two methods, 
linear regression (``T--T plot'') and maximum likelihood. We partition 
the sky into 24 disjoint sky regions, and evaluate the spectral index 
for all polarization angles between $0^{\circ}$ and $85^{\circ}$ in steps of 
$5^{\circ}$. Averaging over polarization angles, we derive a mean 
spectral index of $\beta^{\textrm{all-sky}} = -2.99\pm0.01$ 
in the frequency range of 23-33 GHz. We find 
that the synchrotron spectral index steepens by $0.14$ from low 
to high Galactic latitudes, in agreement with previous studies, with 
mean spectral indices of $\beta^{\textrm{plane}} = -2.98\pm0.01$ and 
$\beta^{\textrm{high-lat}} = -3.12\pm0.04$. In addition, 
we find a significant longitudinal variation along 
the Galactic plane with a steeper spectral index toward
the Galactic center and anticenter than toward the Galactic spiral
arms. This can be well modeled by an offset sinusoidal, 
$\beta(l) = -2.85 + 0.17\sin(2l - 90^{\circ})$. Finally, we study 
synchrotron emission in the BICEP2 field, in an attempt to 
understand whether the claimed detection of large-scale B-mode 
polarization could be explained in terms of synchrotron contamination. 
Adopting a spectral index of $\beta=-3.12$, typical for high Galactic
latitudes, we find that the most likely bias corresponds to about 2\%
of the reported signal ($r=0.003$). The flattest index allowed by the 
data in this region is $\beta=-2.5$, and under the assumption of a 
straight power-law frequency spectrum, we find that synchrotron 
emission can account for at most 20\% of the reported BICEP2 signal.

\end{abstract}

\section{Introduction}
\label{sec:introduction}

Increasingly detailed observations of the cosmic microwave background
(CMB) have revolutionized cosmology during the last 2
decades. Through experiments such as COBE \citep{mather:1990}, the
\emph{Wilkinson Microwave Anisotropy Probe} (\emph{WMAP})
\citep{bennett:2013} and \emph{Planck} \citep{planckI:2014}, not to mention a
host of ground-based and suborbital experiments, a cosmological
concordance model has been established. With only a handful of free
parameters, this model is able to fit literally millions of observed
data points \citep[e.g.,][]{planckXVI:2014}.

These observations have led not only to a cosmological revolution, but
also to a dramatic improvement of our knowledge of the Milky Way. Two
recent and powerful examples of this are the thermal dust and CO maps
published by \emph{Planck}, providing a detailed picture of two individual
Galactic components at an angular resolution of 5 and 10 arcmin,
respectively \citep{planckXI:2014,planckXII:2014,planckXIII:2014}. 

The key to deriving astrophysical component maps lies in the frequency
spectrum of the observed sky: since each physical emission process
results in a different frequency spectrum, in general it is possible
to fit some effective parametric signal model to a set of
multifrequency observations \citep[e.g.,][]{eriksen:2006,eriksen:2008}. A wide
range of methods that performs this inversion has already been
proposed in the literature, and the underlying methodology is well
established by now \citep[see, e.g.,][and references
  therein]{planckXII:2014}.

The main outstanding problem in CMB component separation today is thus
not algorithmic, but rather one of data starvation. For instance, we
know today that there are at least four different significant
temperature emission processes between, say, 20 and 70 GHz, namely
CMB, synchrotron, free-free and, most likely, spinning dust emission
\citep[e.g.,][]{bennett:2013,planckXII:2014}. The minimum number of
parameters required to model this system is therefore seven,
allowing for at least one spectral parameter per foreground
component. This is precisely the same number of frequency channels
that is available in the same frequency range when combining \emph{WMAP}
and \emph{Planck}. In other words, the system is intrinsically nonrigid and
almost degenerate with currently available data.

To make further progress on resolving these components, it is
essential to fully exploit all pieces of available information. One
direction is to use auxiliary data taken at non-CMB frequencies, such
as the 408, 1420 and 2300 MHz maps observed by \citet{haslam:1982},
\citet{reich:1982} and \citet{carretti:2013} or using H$\alpha$ data 
\citep[e.g.,][]{dickinson:2003}. A second direction is to
exploit polarization information: since both free-free and spinning
dust emission are expected to be only weakly polarized 
\citep[see, e.g.,][for observations]{macellari:2011,dickinson:2011}
\citep[see, e.g.,][for theory]{hoang:2013}, there is only
one known significant foreground emission mechanism at low CMB
frequencies ($\sim10-70$ GHz), namely synchrotron emission,
which is caused by relativistic electrons spiraling in 
the Galactic magnetic field. Models of the Galactic B field 
\citep[e.g.,][]{fauvet:2012} and the energy distribution of the cosmic ray 
electrons \citep[e.g.,][]{orlandostrong:2013} can then be used to model 
synchrotron radiation in programs like the 
GALPROP\footnote{http://galprop.stanford.edu/} code.

In this paper, we consider the \emph{WMAP} K (23 GHz) and Ka (33 GHz) bands 
and measure the effective spectral index between these in various regions 
on the sky. The result is a map of the spectral index of polarized
synchrotron emission that for instance may be used as a prior to
inform more advanced and complete analyses. This map also represents
an important result in its own right, since the specific value of the
spectral index carries information about the physical conditions at
the emission origin.

A similar analysis was carried out for the 5 yr \emph{WMAP} data by
\citet{dunkley:2009}, who used a Gibbs sampling technique to measure
the synchrotron spectral index over a low-resolution grid with
$30^{\circ}\times30^{\circ}$ pixels. In the present paper, we employ
two different algorithms to the same goal in order to understand the
robustness of the particular method of choice. The first method is
simple linear regression as implemented in a so-called ``T--T plot''
technique \citep{turtle:1962}, which is well known in the radio
astronomy literature, and enjoys significant popularity due to its
insensitivity to arbitrary offsets in the data. The second method is a
standard maximum likelihood (ML) method, which in principle is similar to
the Gibbs sampler employed by \citet{dunkley:2009}. However, there are
at least four important differences between these two analyses. First,
we marginalize over unknown offsets within each region, to ensure the
same robustness in the likelihood approach as in the ``T--T plot''
technique. Second, we consider data smoothed to $1^{\circ}$ FWHM
whereas \citet{dunkley:2009} considered data downgraded to
$4^{\circ}\times4^{\circ}$ pixels. Third, we define a set of 24
physically motivated regions, whereas \citet{dunkley:2009} adopted a
regular grid of 48 $30^{\circ}\times30^{\circ}$ low-resolution
pixels. Finally, we study the 9 yr \emph{WMAP} observations, whereas
\citet{dunkley:2009} analyzed the 5 yr \emph{WMAP} observations.
This longer period of data taking is especially important 
in regions at high Galactic latitude where the signal-to-noise ratio 
is low.

The importance of this topic was highlighted with the recent release
of the new BICEP2 large-scale polarization observations in 2014 March
\citep{bicep2:2014}. Based on these measurements, the team claimed the
first detection of primordial B modes with an amplitude corresponding
to a tensor-to-scalar ratio of $r=0.20^{+0.07}_{-0.05}$, formally
ruling out the hypothesis of no B-mode signal beyond gravitational
lensing at $7\,\sigma$. If confirmed, and shown to be cosmological, this
claim will have fundamental consequences for cosmology. One part of
that validation process is to understand whether any astrophysical
signals, for instance polarized synchrotron radiation emission, could
explain part of the excess. We address this question at the end of the
paper.

\section{Methods}
\label{sec:method}

\subsection{Spectral indices by T--T Plots}
\label{sec:ttplots}

We start by reviewing the linear regression, or ``T--T plot'',
technique \citep{turtle:1962}. Let us first assume that we have
observational polarization data in the form of two images of some
extended region with an intrinsically constant spectral index, but
with spatially varying amplitudes across the field, taken at different
frequencies,
\begin{equation}
\mathbf{d}_{\nu} = \mathbf{A} \left(\frac{\nu}{\nu_0}\right)^{\beta} +
\mathbf{n}_{\nu}. 
\label{eq:model0}
\end{equation}
Here $\mathbf{d}_{\nu}$ denotes a vector of the Stokes $Q$ and $U$
parameters at frequency $\nu$ and pixel $p$, $\nu_0$ is some arbitrary
reference frequency, $\mathbf{A}$ is the amplitude of the signal at
$\nu_0$, $\beta$ is the spectral index we seek to determine, and
$\mathbf{n}_{\nu}$ denotes instrumental noise, which is typically
assumed Gaussian with zero mean and known covariance, $\mathbf{N}$. 

\begin{figure}[t]
\begin{center}
\mbox{\epsfig{figure=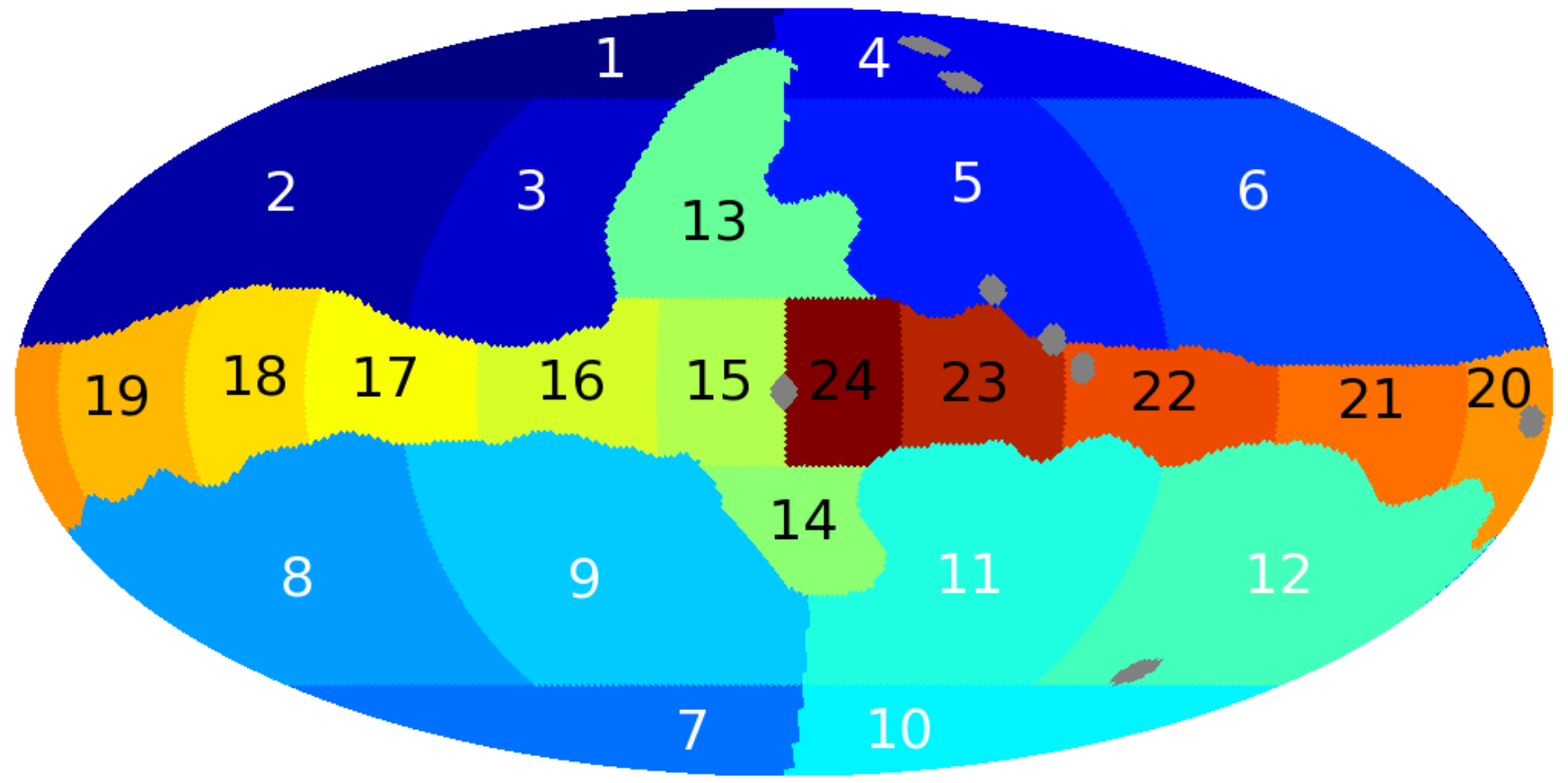,width=\linewidth,clip=}}
\mbox{\epsfig{figure=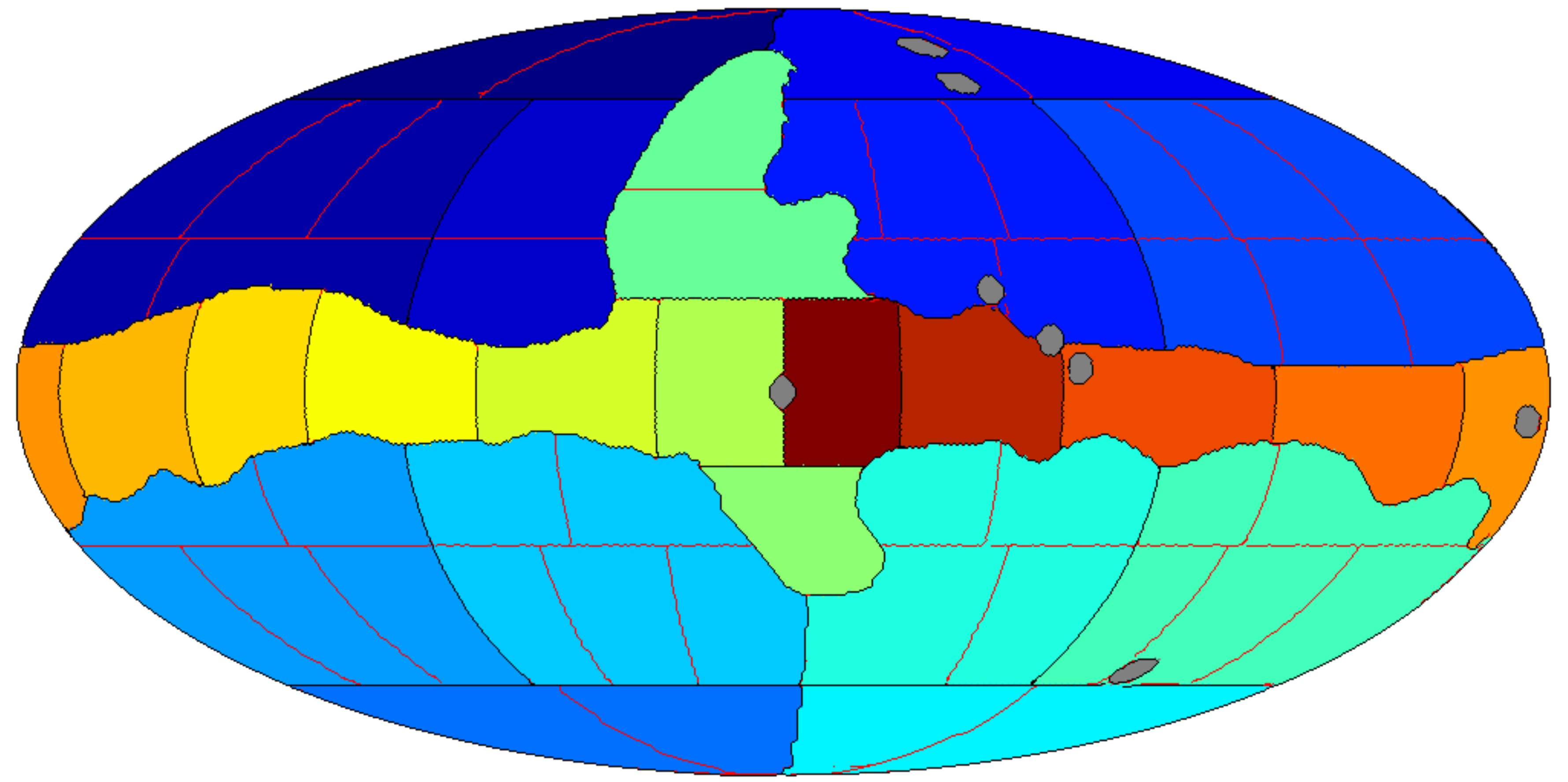,width=\linewidth,clip=}}
\end{center}
\caption{Top: main region definition adopted for this analysis.
  The sky is divided into 24 regions, removing particularly bright
  point sources and the Galactic center.
  Bottom: the large high-latitude regions are divided into 
  subregions for use with the offset determination in the maximum 
  likelihood (ML) method.}
\label{fig:regionmask}
\end{figure}

For the ideal and noiseless case, $\mathbf{n}_{\nu}=0$, we see from
Equation~\eqref{eq:model0} that the spectral index can be found simply
as the ratio of the amplitudes at each pixel, weighted by the
frequency lever arm,
\begin{equation}
\label{eq:tt_ideal}
\frac{d_{\nu_1}(p)}{d_{\nu_2}(p)} = \left(\frac{\nu_1}{\nu_2}\right)^{\beta}
\Rightarrow \beta = \frac{\log (d_{\nu_1}(p)/d_{\nu_2}(p))}{\log (\nu_1/\nu_2)}.
\end{equation}
Thus, the data at the first frequency depends linearly on the data at
the second frequency, $d_1 = ad_2 + b$, with a slope given by
$a=(\nu_1/\nu_2)^{\beta}$, and for the ideal case, the intercept is
zero, $b=0$. However, note that any constant offset in either $d_1$ or
$d_2$ translates directly into a nonzero value of $b$, but does
not change the slope. Thus, the spectral index, as estimated by
this technique, is fully insensitive to spurious offsets in the data,
and this is the primary reason for the popularity of the method.

In practice, data are never perfect or noiseless, and the above
relation therefore only holds statistically. Instead, our data set
consists of $N$ frequency data pairs, $\{d_{\nu_1}(p),
d_{\nu_2}(p)\}$, to which we can fit a straight line.  One method for
doing this is through a standard least-squares fit. However, it is
important to note that the data in this case typically have
uncertainties in both $d_1$ and $d_2$ directions, and the standard
textbook least-squares algorithm is in this case biased. An equivalent
method with support for noise in both directions is the effective
variance method \citep{orear:1982,petrolini:2011}, with an error
function on the form
\begin{equation}
S(a,b) = \sum_{p} \frac{(d_{1}(p) - ad_{2}(p) -b)^2}{\sigma_1^2+(\partial d_1(p)/ \partial d_2(p))^2 \sigma_2^2}.
\end{equation}
Assuming for simplicity that the errors in the two directions are the
same, which is a very good approximation for the \emph{WMAP} observations, we
can minimize this function by equating the partial derivatives with
zero,
\begin{equation}
a = D + \frac{C}{|C|} \sqrt{1+D^2}
\end{equation}
\begin{equation}
b = \langle d_1 \rangle - a \langle d_2 \rangle
\end{equation}
\begin{equation}
D = \frac{V_1 - V_2}{2C}
\end{equation}
\begin{equation}
V_2 = \langle d_2^2 \rangle - \langle d_2 \rangle^2  \qquad  V_2 = \langle d_1^2 \rangle - \langle d_1 \rangle^2
\end{equation}
\begin{equation}
C = \langle d_1d_2 \rangle - \langle d_1 \rangle \langle d_2 \rangle.
\end{equation}
The spectral index, $\beta$, is then
\begin{equation}
\label{eq:tt_noise}
a = \left(\frac{\nu_1}{\nu_2}\right)^{\beta}
\Rightarrow \beta = \frac{\log a}{\log (\nu_1/\nu_2)}.
\end{equation}
The error in the slope $a$ reads
\begin{equation}
\sigma_{a}=(1+a^2) \sqrt{\frac{1}{N-2} \frac{V_1+V_2}{(V_1-V_2)^2 + 4C^2} },
\end{equation}
and using the relation between standard errors, $\sigma_{\beta}=(\frac{d \beta}{d a}) \sigma_a $,
the error in the spectral index is
\begin{equation}
\sigma_{\beta} = \frac{\sigma_a}{a} \frac{1}{\log (\nu_1/\nu_2)}.
\end{equation}
Note that this is only a statistical error. In order to get a more
realistic error estimate we need to add a systematic uncertainty term.
In this paper we estimate this via bootstrap sampling: We randomly
draw 10\,000 new data sets from the original data set, each consisting
of $N$ pairs of data points, and duplicate points are allowed. The
analysis is then done on each subsample, each resulting in one value
of the spectral index, and the resulting standard deviation is adopted
as the systematic error.

\begin{figure*}[t]
\begin{center}
\mbox{\epsfig{figure=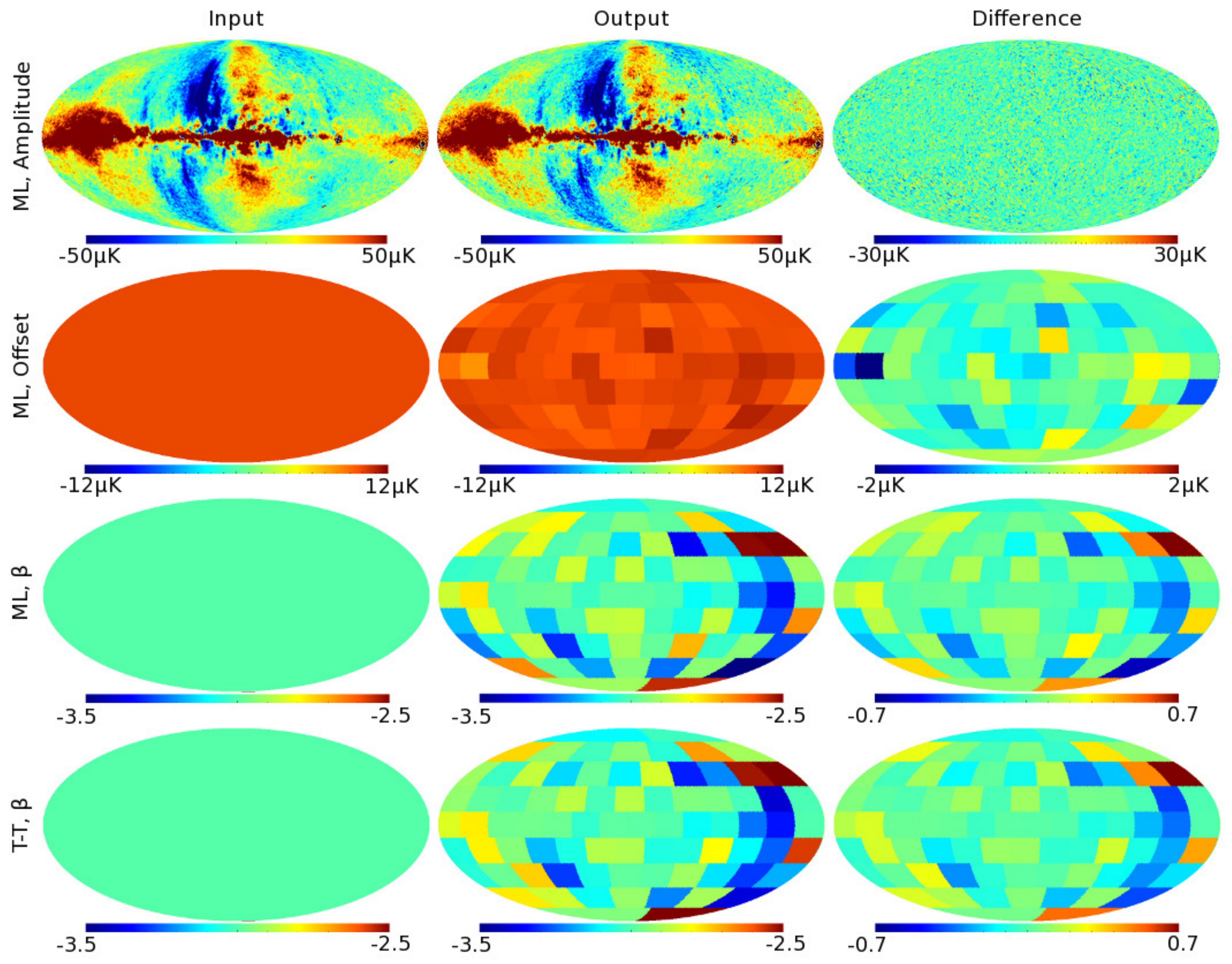,width=\linewidth,clip=}}
\end{center}
\caption{Validation by simulations.  The columns show, from left to
  right, (1) the true input sky maps, (2) the derived sky maps, and (3) the
  difference between the two. Rows show, from top to bottom, (1) the
  Stokes $Q$ amplitude, (2) the $Q$ offset and (3) spectral index,
  $\beta_{\textrm{tot}}$, for the ML method, and
  (4) the spectral index, $\beta_{\textrm{tot}}$, for the T--T plot technique.}
\label{fig:simulation}
\end{figure*}

\begin{figure}[t]
\begin{center}
\mbox{\epsfig{figure=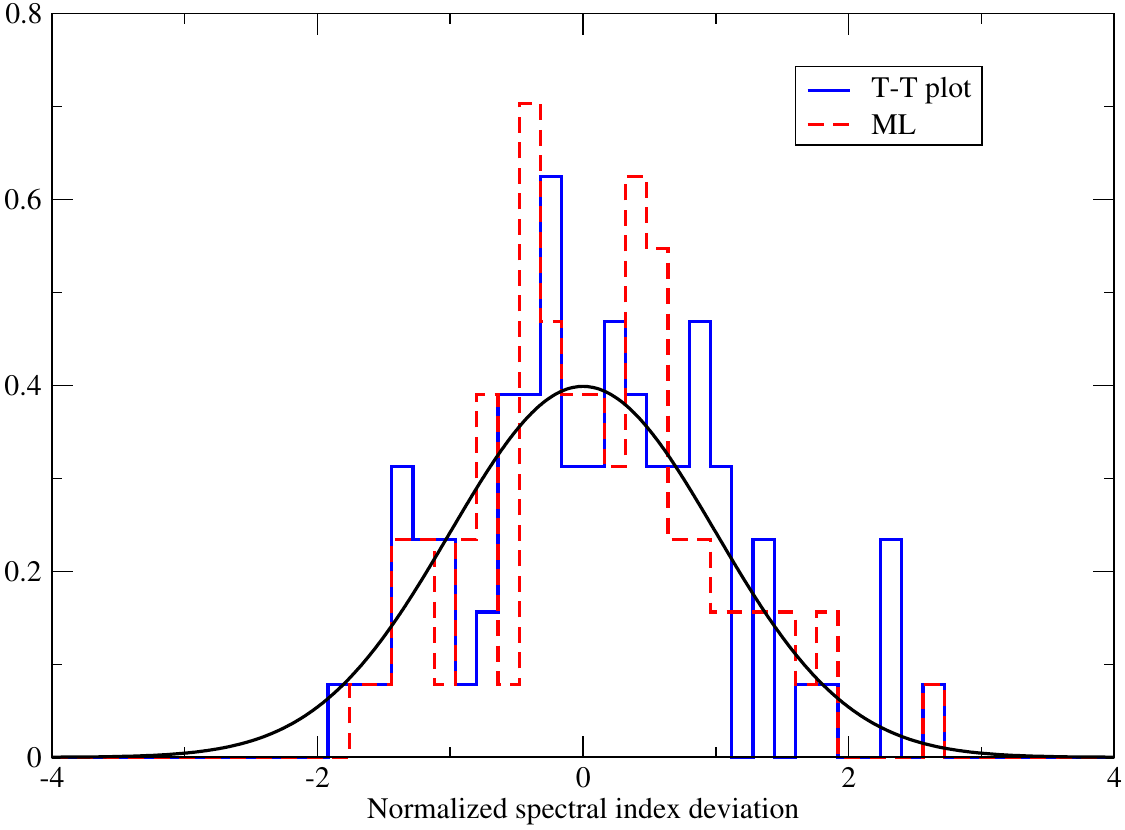,width=\linewidth,clip=}}
\end{center}
\caption{Histogram of normalized spectral index deviations, 
$(\beta_{tot} - \beta_{\textrm{sim}})$/$ \sigma_\beta$, for T--T plot method (blue) 
and maximum likelihood method (red, dashed) for the simulation.
The black curve shows a standard normal distribution with zero mean
and unit variance.}
\label{fig:simhistogram}
\end{figure}

\subsection{Basic maximum likelihood estimation of spectral indices}
\label{sec:ml_analysis}

The main advantage of the T--T plot method is implementational
robustness, by virtue of being fully insensitive to absolute
offsets. However it is neither very extendable nor does it easily
provide well-defined uncertainties. For these reasons we want to
define a ML method that provides similar robustness as
the T--T plot approach, but still expresses the full uncertainties in
terms of a proper probability distribution. To do so, we extend the
data model in Equation \eqref{eq:model0} with an offset map,
$\mathbf{m}_{\nu}$, at each frequency,
\begin{equation}
\mathbf{d}_{\nu} = \mathbf{A} \left(\frac{\nu}{\nu_0}\right)^{\beta} +
\mathbf{m}_{\nu} + \mathbf{n}_{\nu}.
\label{eq:model1}
\end{equation}
Since the noise is assumed Gaussian with covariance $\mathbf{N}$, it
is straightforward to write down the likelihood for this model,
\begin{equation}
\label{eq:likelihood}
\begin{split}
-2\log\mathcal{L}(\mathbf{A}, \mathbf{m}_\nu, \beta) \propto& \\\sum_{\nu}
\left(\mathbf{d}_\nu - \mathbf{A}(\nu/\nu_0)^{\beta} -
\mathbf{m}_\nu \right)^t &\mathbf{N}^{-1}\left(\mathbf{d}_\nu -
\mathbf{A}(\nu/\nu_0)^{\beta} - \mathbf{m}_\nu
\right).
\end{split}
\end{equation}

If we define the offset map as spatially constant, this approach
retains the exact same degrees of freedom as the T--T plot
method. However, contrary to the T--T method, this framework also
allows subdivision of the offset map into smaller regions, thereby
trading signal-to-noise against the ability to trace large-scale
features, for instance due to correlated noise. In this paper, we
divide the largest regions into subregions 
as specified in Section \ref{sec:data}.

With this data model and likelihood, the optimal likelihood estimate
for $\{\mathbf{A}, \mathbf{m}_{\nu}, \beta\}$ may now be determined,
for instance using a standard Newton-Raphson optimizer or Powell's 
search method or a Gibbs sampler or even a simple grid evaluation, 
with corresponding uncertainty estimates given either by Fisher 
matrix approximations or proper marginals. In this paper, we adopt 
Powell's method with Fisher matrix approximations.

\subsection{Marginalizing over polarization angle}
\label{sec:practicalities}

As shown by \citet{wehus:2013}, the synchrotron spectral index from the
\emph{WMAP} K- and Ka-band observations is not stable with respect to polarization
orientation even for a supposedly stable source such as Tau-A.  To
obtain robust results we therefore marginalize over the polarization
angle. Specifically, we first rotate the data by a set of angles,
$\alpha$, into new coordinate systems, $\mathbf{d}(\alpha)=Q
\cos2\alpha + U \sin2\alpha$, letting $\alpha$ vary between $0^\circ$
and $85^\circ$ in steps of $5^\circ$; $\alpha=0^\circ$ and
$\alpha=45^\circ$ correspond to measuring the spectral index from
Stokes $Q$ or $U$ only.  After this operation, we have 18 (highly
dependent) data sets, from which we compute an average spectral index
by inverse-variance weighting,
\begin{equation}
\beta_{\textrm{tot}} = \frac{\sum_{i=1}^{18} \beta_i / \sigma_i^2}{\sum_{i=1}^{18} 1 / \sigma_i^2}.
\end{equation}
Attaching a sensible uncertainty to this estimate is difficult, as
systematic errors from, e.g., beam ellipticities are not
negligible. For now, we simply adopt the minimum of the individual
uncertainties as the error estimate, noting that adding more
observations should never increase the statistical uncertainties.

\begin{figure*}[p]
\begin{center}
\mbox{\epsfig{figure=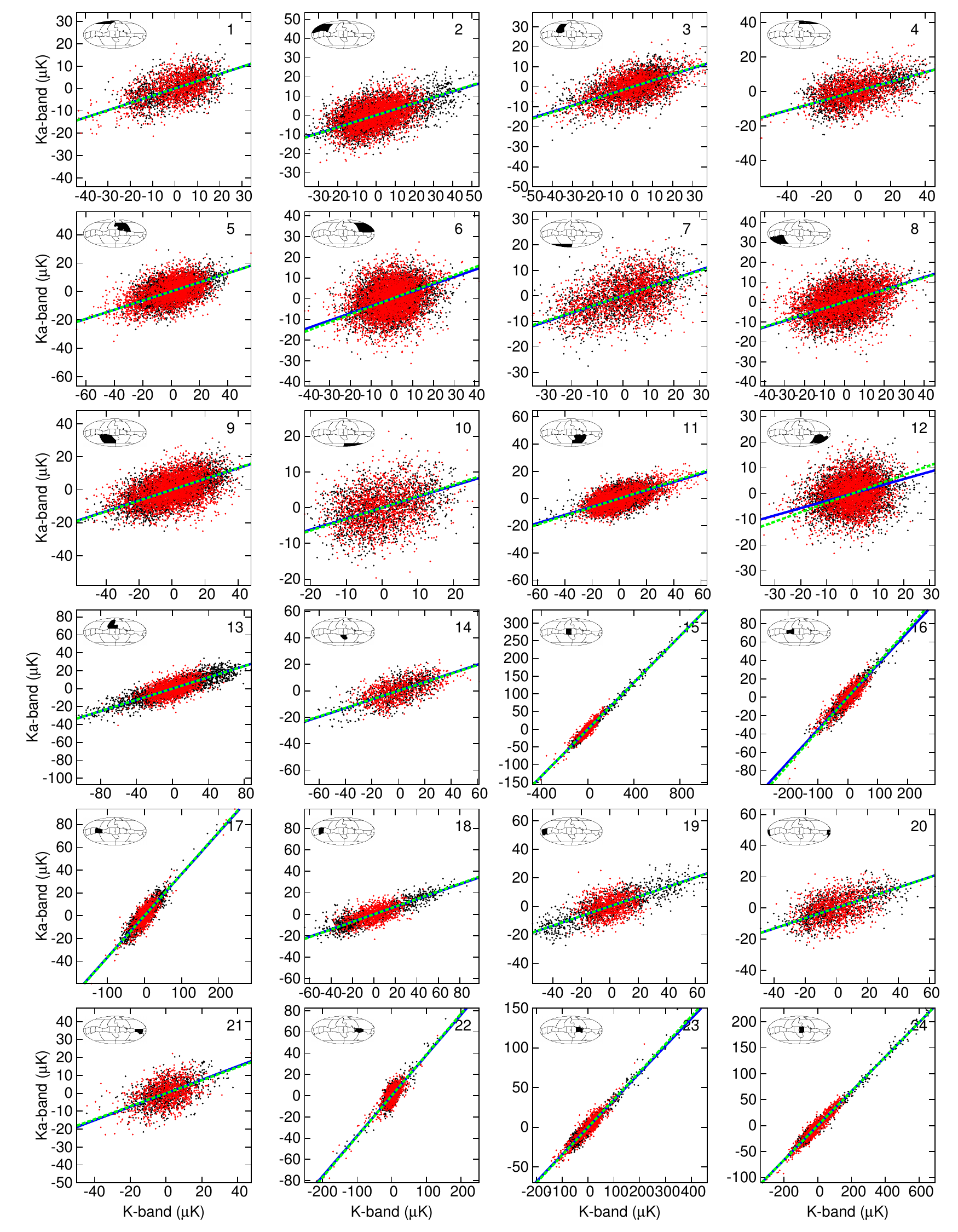,width=\linewidth,clip=}}
\end{center}
\caption{T--T plots for Stokes $Q$ (black) and $U$ (red) in regions 
  1-24.  The two lines correspond to the best-fit spectral indices
  derived with the T--T plot (solid blue) and maximum likelihood
  methods (green, dashed).}
\label{fig:scatterplots}
\end{figure*}

\begin{figure*}[p]
\begin{center}
\mbox{\epsfig{figure=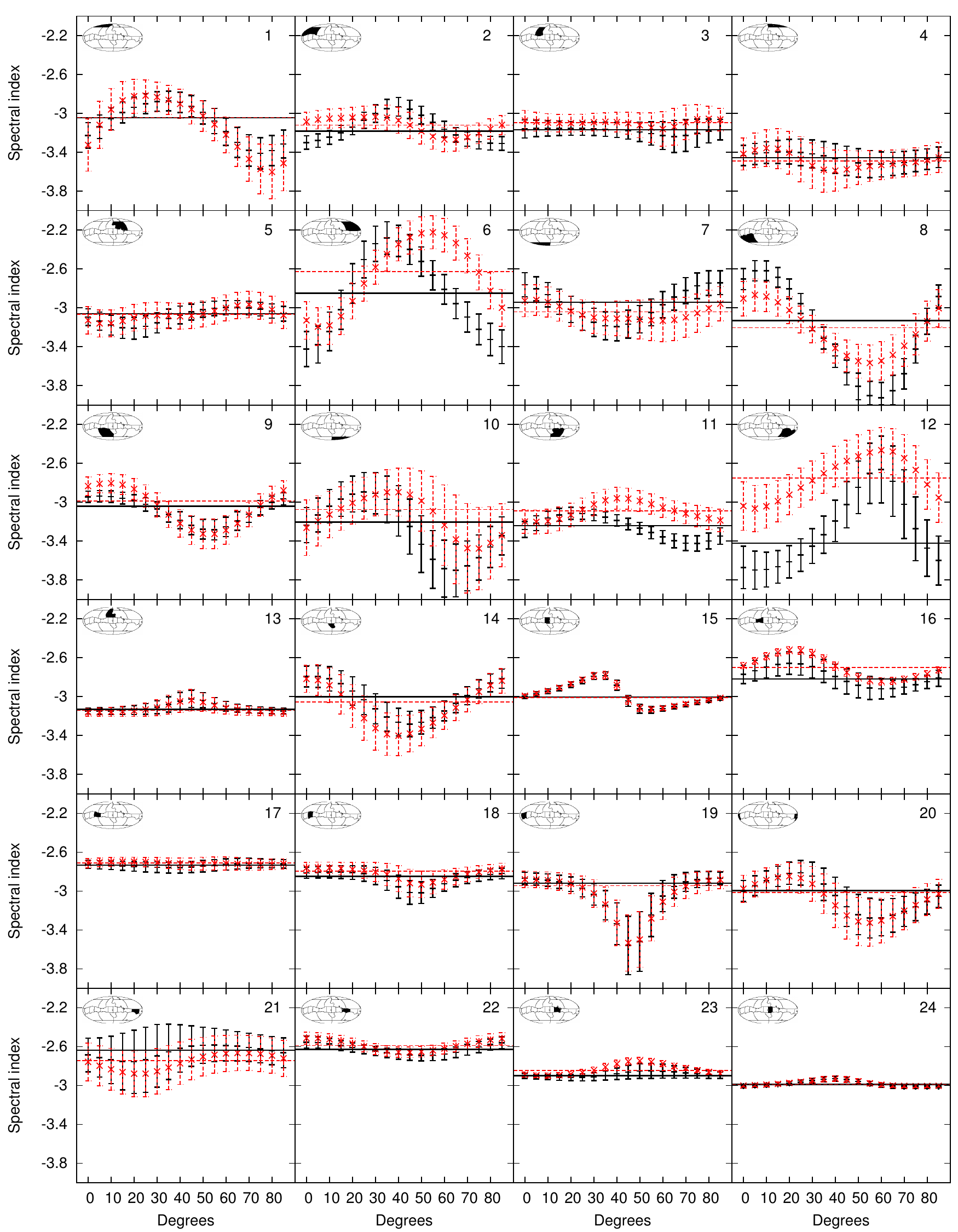,width=\linewidth,clip=}}
\end{center}
\caption{Spectral index as a function of polarization orientation for
  T--T plot (black) and maximum likelihood (red, dashed).}
\label{fig:alphaplots}
\end{figure*}

\section{Data}
\label{sec:data}

The main goal of this paper is to measure the spectral index of
polarized synchrotron emission from the 9 yr \emph{WMAP} polarization
data.\footnote{http://lambda.gsfc.nasa.gov} We therefore focus on the
two lowest frequencies, the K and Ka bands with effective frequencies
of 22.45 and 32.64 GHz, respectively, for a synchrotron spectral index
of $\beta=-3$ \citep{page:2003}. At the Ka band, the typical 
level of CMB and dust emission is 1\%-2\% that of synchrotron, and 
at the K band it is several times smaller. This implies that both channels 
are strongly synchrotron dominated on the scales of interest in this 
paper, and we therefore neglect both thermal dust emission and CMB 
fluctuations in the following.

The \emph{WMAP} K and Ka bands have angular resolutions of 53 and 40 arcmin
FWHM, respectively, and are pixelized at a
HEALPix\footnote{http://healpix.jpl.nasa.gov} resolution of
$N_{\textrm{side}}=512$ (6.7 arcmin). In our analyses we require the 
data to be at a common resolution and therefore smooth both maps to 
a common resolution of $1^{\circ}$ FWHM and rebin them onto an
$N_{\textrm{side}}=64$ (55 arcmin) HEALPix grid. 

Although observing at relatively low frequencies, the \emph{WMAP}
polarization maps are strongly noise dominated at high Galactic
latitudes. To achieve a reasonable signal-to-noise ratio over most of
the sky, we therefore partition the sky into 24 disjoint regions,
shown in the top panel of Figure \ref{fig:regionmask}. 
The starting point of the region definitions is the
P06 polarization mask provided with the \emph{WMAP} data, smoothed with a 
median filter. Inside this mask we expect the polarization 
foregrounds to be dominating, and we therefore construct a set of 
smaller regions inside the mask and larger regions outside. In addition,
the offsets in the ML method are defined by subdividing the large
high-latitude regions according to Galactic latitude and longitude,
such that each subregion contains typically $\sim1000$ pixels, as 
shown in the bottom panel of Figure \ref{fig:regionmask}. Particularly 
bright compact objects are excluded from the analysis, as is the 
Galactic center.

Because of the high pixel resolution of the \emph{WMAP} sky maps, only
per-pixel noise characterization is provided by the \emph{WMAP} team for the
full-resolution maps. (Correlated pixel noise
covariance matrices are only in $N_{\textrm{side}}=16$.) However, 
from Table 2 in \citet{jarosik:2003} 
we see that the $f_{\textrm{knee}}$ values for the \emph{WMAP} K and Ka band 
radiometers range from 0.3 to 0.7 mHz, and the noise may therefore 
be approximated as white.
The noise maps are given in the form of $2\times2$ Stokes $Q$, $U$
submatrices. From these, we generate full pixel-pixel noise
covariance matrices for each (sub)region separately, accounting for
the smoothing operation that has been applied to the maps. These
matrices are subsequently propagated into the likelihood analyses.

\section{Validation by simulations}
\label{sec:simulations}

Before applying our methods to the real data, we analyze simulations
for validation purposes.  These simulations are generated by adopting
the (smoothed) \emph{WMAP} K-band map as a perfect synchrotron template at
K band, to which noise is added according to the \emph{WMAP} noise model. To
generate the corresponding Ka-band channel, we scale the template to
32.64 GHz assuming a spectral index of $\beta=-3$; adopting a single
spectral index for all regions makes it easy to spot visually outliers
and errors in the resulting maps. Finally, we add an offset of
$\mathbf{m}_{\textrm{Ka}}=10\,\mu\textrm{K}$ to the Ka band. For this
initial test, we adopt a sky tessellation consisting of latitudinal and
longitudinal squares as our region definitions.

These simulations are processed using both the ML and
the T--T plot techniques described in Section \ref{sec:method}, and
the main results are summarized in Figure
\ref{fig:simulation}. Columns show, from left to right, input, output
and difference maps, and rows show, from top to bottom, the Stokes $Q$
amplitude, $Q$ offset and spectral index as computed with the ML
method, and, finally, the spectral index as computed with the T--T
method. For the spectral index maps, we find an (inverse-variance
weighted) average of $\beta= -2.996\pm 0.005$ for T--T plot and
$\beta= -2.995\pm 0.007$ for ML.

In Figure \ref{fig:simhistogram} we plot histograms for the normalized
spectral index deviations, $(\beta_{\textrm{tot}} -
\beta_{\textrm{sim}})$/$ \sigma_\beta$ for the ML and T--T methods. If
our methods are both unbiased and produce sensible uncertainty
estimates, these should match a standard normal distribution,
$N(0,1)$, indicated by a solid black line. The standard deviations of
the two histograms are 0.87 and 0.98 for the ML and T--T plot 
techniques, respectively, indicating that both methods perform well.

\section{All-sky analysis}
\label{sec:results}

We now turn to the actual 9 yr \emph{WMAP} K- and Ka-band polarization
data and show first in Figure \ref{fig:scatterplots} T--T scatter
plots for each of our predefined 24 regions. Black and red dots show
Stokes $Q$ and $U$ parameters, respectively; adopting different
coordinate systems correspond to linear combinations between these
distributions. The lines indicate the best-fit spectral indices
obtained by the T--T plot (solid blue) and ML (green, dashed)
methods. The different synchrotron signal-to-noise ratios from region
to region are clearly seen here as different scatter plot
ellipticities; regions with a high signal-to-noise ratio have scatter
plots that are highly elongated, whereas regions with low signal-to-noise
ratios are almost circular.

\begin{deluxetable}{rccc}
\tablewidth{0pt}
\tablecaption{\label{tab:spectralindex}Synchrotron
  spectral index for each region}
\tablecomments{Synchrotron spectral index derived from the 9-year \emph{WMAP}
  polarization data with the maximum likelihood (second column) and
  T--T plot (third column) methods. The algorithm averaged results are
  listed in the fourth column.}
\tablecolumns{4}
\tablehead{Region  &     ML       &    T--T plot           &       Combined  } 
\startdata
1 & $  -3.04\pm 0.15 $ & $ -3.05 \pm 0.10 $ & $ -3.04 \pm 0.15 $  \\
2 & $  -3.12\pm 0.10 $ & $ -3.18 \pm 0.07 $ & $ -3.15 \pm 0.16 $  \\
3 & $  -3.09\pm 0.09 $ & $ -3.17 \pm 0.07 $ & $ -3.13 \pm 0.16 $  \\
4 & $  -3.49\pm 0.14 $ & $ -3.46 \pm 0.10 $ & $ -3.47 \pm 0.18 $  \\
5 & $  -3.07\pm 0.13 $ & $ -3.06 \pm 0.07 $ & $ -3.07 \pm 0.14 $  \\
6 & $  -2.63\pm 0.17 $ & $ -2.85 \pm 0.15 $ & $ -2.74 \pm 0.39 $  \\
7 & $  -3.04\pm 0.15 $ & $ -2.94 \pm 0.12 $ & $ -2.99 \pm 0.25 $  \\
8 & $  -3.20\pm 0.14 $ & $ -3.13 \pm 0.10 $ & $ -3.17 \pm 0.21 $  \\
9 & $  -2.99\pm 0.10 $ & $ -3.04 \pm 0.06 $ & $ -3.01 \pm 0.15 $  \\
10 & $  -3.07\pm 0.22 $ & $ -3.20 \pm 0.17 $ & $ -3.14 \pm 0.34 $  \\
11 & $  -3.09\pm 0.10 $ & $ -3.24 \pm 0.06 $ & $ -3.16 \pm 0.26 $  \\
12 & $  -2.75\pm 0.20 $ & $ -3.42 \pm 0.17 $ & $ -3.09 \pm 0.87 $  \\
13 & $  -3.14\pm 0.04 $ & $ -3.13 \pm 0.03 $ & $ -3.14 \pm 0.05 $  \\
14 & $  -3.05\pm 0.12 $ & $ -3.00 \pm 0.09 $ & $ -3.03 \pm 0.18 $  \\
15 & $  -3.01\pm 0.01 $ & $ -3.00 \pm 0.02 $ & $ -3.01 \pm 0.02 $  \\
16 & $  -2.70\pm 0.03 $ & $ -2.82 \pm 0.06 $ & $ -2.76 \pm 0.15 $  \\
17 & $  -2.71\pm 0.05 $ & $ -2.73 \pm 0.04 $ & $ -2.72 \pm 0.07 $  \\
18 & $  -2.79\pm 0.06 $ & $ -2.85 \pm 0.05 $ & $ -2.82 \pm 0.11 $  \\
19 & $  -2.94\pm 0.08 $ & $ -2.92 \pm 0.06 $ & $ -2.93 \pm 0.11 $  \\
20 & $  -3.01\pm 0.13 $ & $ -3.00 \pm 0.10 $ & $ -3.00 \pm 0.15 $  \\
21 & $  -2.74\pm 0.18 $ & $ -2.64 \pm 0.14 $ & $ -2.69 \pm 0.29 $  \\
22 & $  -2.59\pm 0.05 $ & $ -2.63 \pm 0.05 $ & $ -2.61 \pm 0.09 $  \\
23 & $  -2.84\pm 0.02 $ & $ -2.90 \pm 0.02 $ & $ -2.87 \pm 0.08 $  \\
\vspace*{1mm}
24 & $  -2.99\pm 0.01 $ & $ -2.99 \pm 0.02 $ & $ -2.99 \pm 0.01 $  \\
Mean & $  -2.96\pm 0.01 $ & $ -2.98 \pm 0.01 $ & $ -2.99 \pm 0.01 $
\enddata
\end{deluxetable}

\begin{figure}[t]
\begin{center}
\mbox{\epsfig{figure=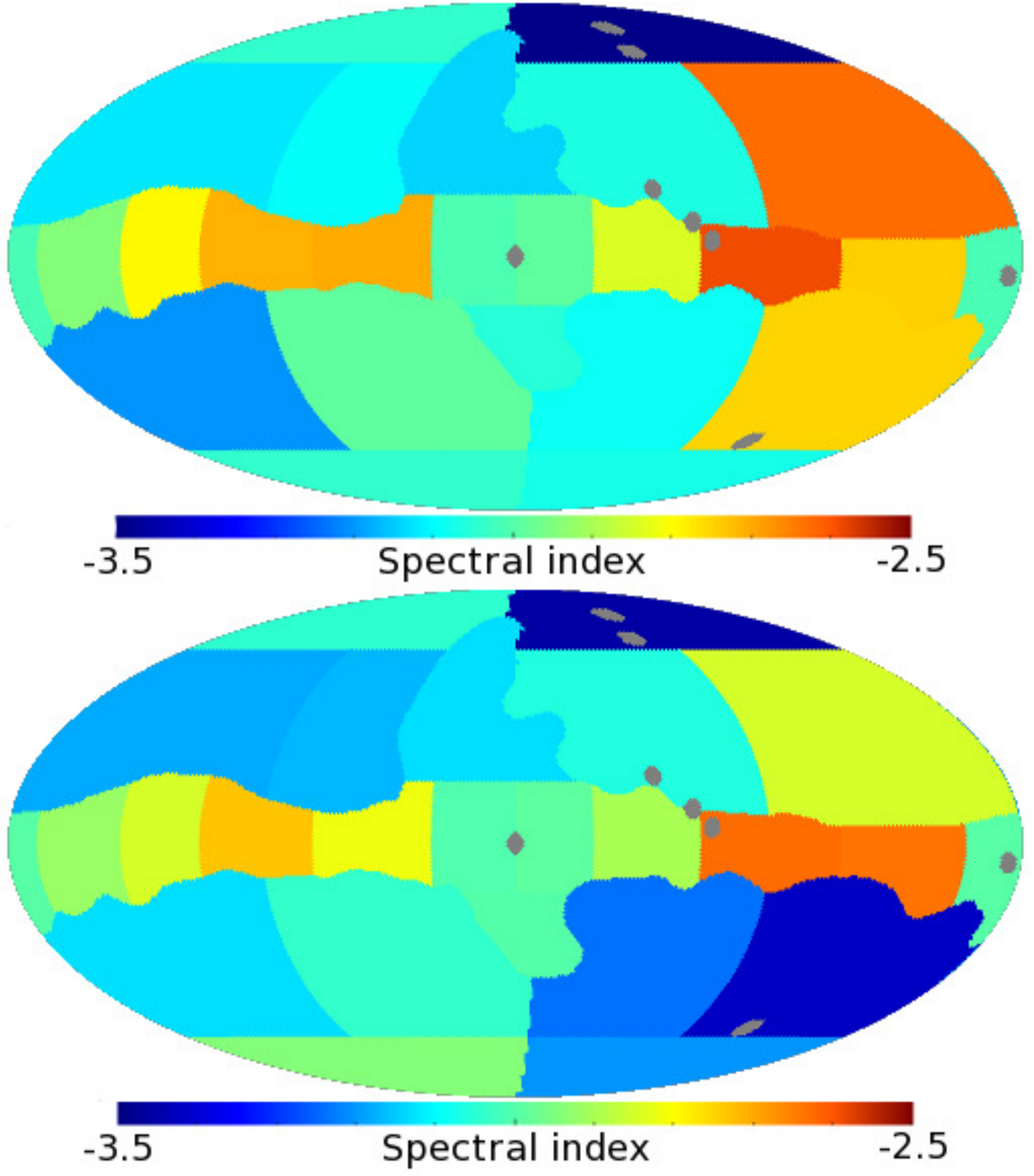,width=\linewidth,clip=}}
\end{center}
\caption{Synchrotron spectral index derived with the maximum
  likelihood (top panel) and T--T plot (bottom panel) methods from the
  9 yr \emph{WMAP} K- and Ka-band polarization sky maps.}
\label{fig:spectralindexwmap}
\end{figure}

\begin{figure}[t]
\begin{center}
\mbox{\epsfig{figure=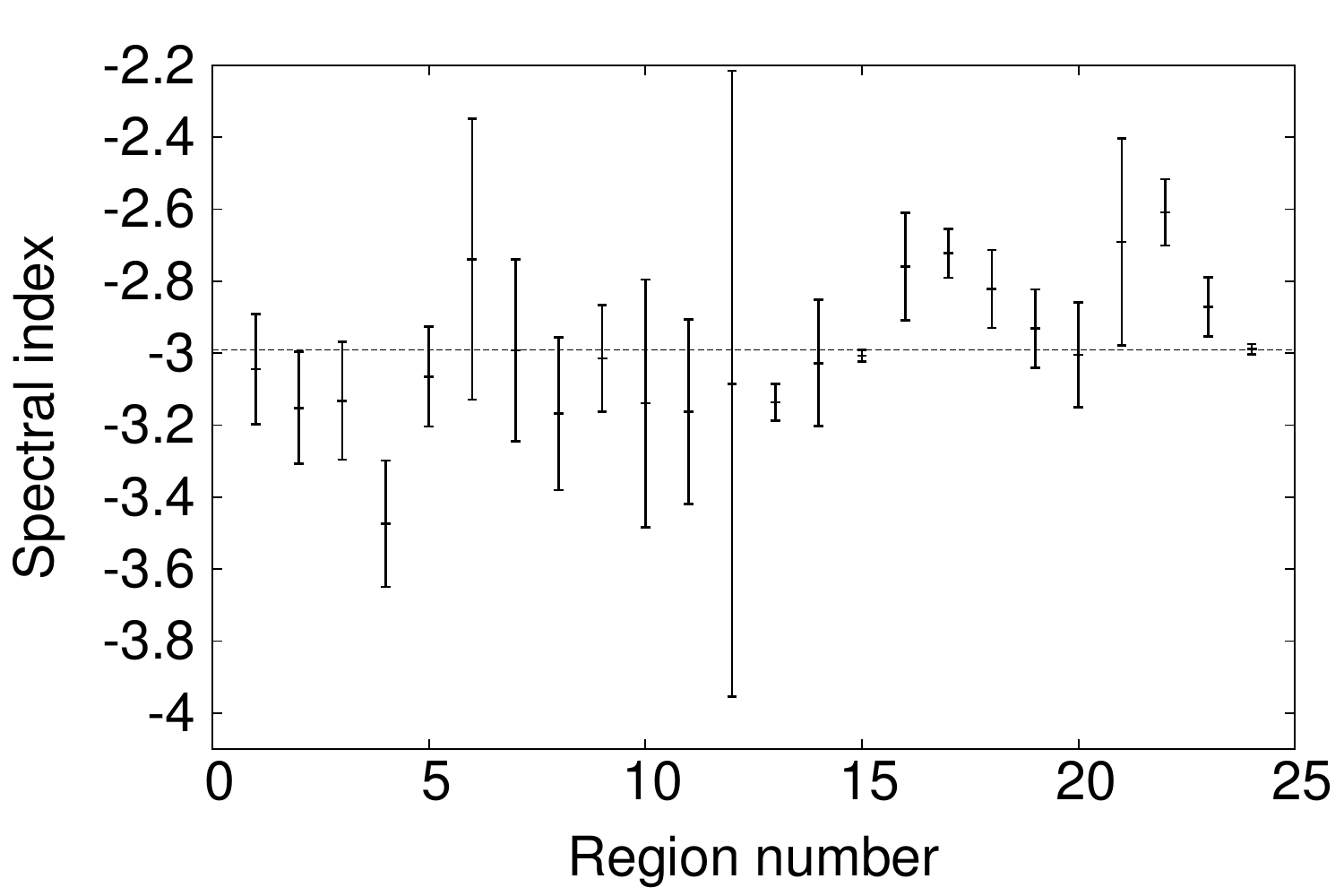,width=\linewidth,clip=}}
\end{center}
\caption{Algorithm-averaged synchrotron spectral index as a function
  of region number. The horizontal dashed line is the inverse-variance weighted
  mean value of all regions, $\beta^{\textrm{all-sky}}=-2.99$.}
\label{fig:spectindexregion}
\end{figure}

\begin{figure}[t]
\begin{center}
\mbox{\epsfig{figure=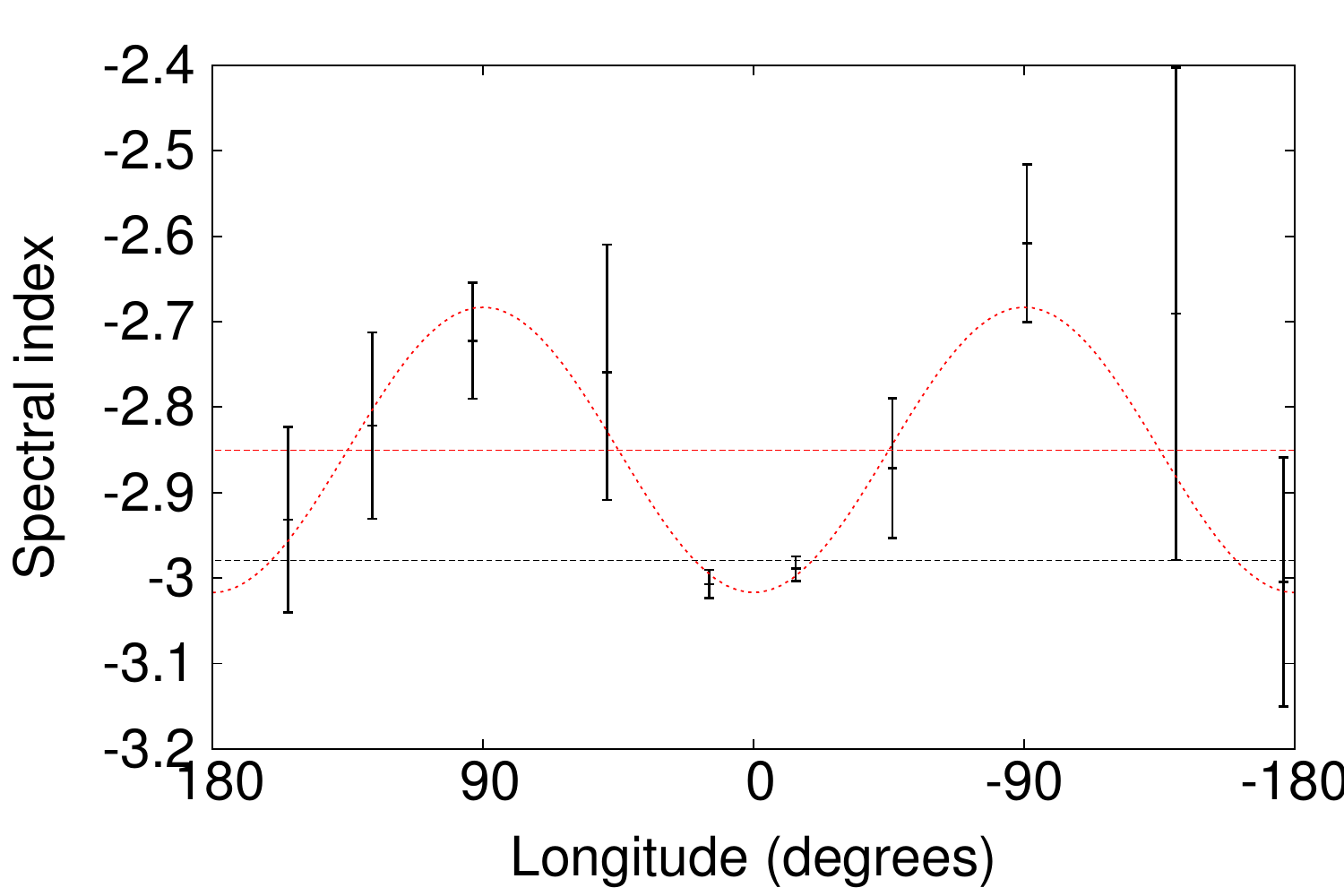,width=\linewidth,clip=}}
\end{center}
\caption{Algorithm-averaged synchrotron spectral index for regions
  along the Galactic plane, plotted as a function of longitude. The
  horizontal dashed black line shows the inverse-variance weighted 
  best-fit constant to these observations, and the red dashed curve 
  shows the best-fit offset sine function, $\beta(l) = c_2 + a\sin(2l -
  90^{\circ})$.  The horizontal red dotted line shows the constant
  $c_2$. Values along the horizontal axis increase from right to
  left, allowing direct mapping onto a Mollweide projection.}
\label{fig:spectindexlongitude}
\end{figure}

In Figure \ref{fig:alphaplots} we plot the derived spectral index as a
function of polarization orientation from $0^\circ$ to $85^\circ$ for
the T--T plot (black) and ML (red) methods. The horizontal lines
indicate the corresponding inverse-variance weighted mean values. In
most regions the agreement between the two methods is good, although
for a few the deviations are substantial. The worst case is region 12,
for which the scatter plot in Figure \ref{fig:scatterplots} is
virtually circular. As a result, the different noise weighting of the
two methods has a large effect. 

In the case of a perfect sky signal with identical spectral index in
both $Q$ and $U$, and contaminated only by noise, the expected
behavior in these plots is that of a simple sinusoidal with period
equal to $45^{\circ}$ and an amplitude given by the random noise
fluctuations in the $Q$ and $U$ parameters. A modulation amplitude
larger than, say, twice the statistical fluctuation, as for instance
is seen in region 15 (close to the Galactic center), therefore either
indicates a true intrinsic variation in the spectral index between the
Stokes $Q$ and $U$ parameters or unmodeled systematics.

The main difference between the T--T and ML methods lies in their
relative noise weighting. While the ML method performs an effective
inverse noise variance weighting, the T--T method weighs all points
equally. One could therefore argue that the ML method is more optimal,
and its results should in principle be more trustworthy. However, we
take a conservative approach and define the observed spectral index
difference between the two methods as an ``algorithmic uncertainty'',
added linearly (as a systematic error) to the statistical
uncertainty. Correspondingly, we adopt the straight mean of the
indices derived with the two methods as our final point estimate of
the spectral index.

Table \ref{tab:spectralindex} lists the derived spectral indices for
all 24 regions for both methods, as well as the combined 
``algorithm-averaged'' values. Figure \ref{fig:spectralindexwmap} 
shows the same in terms of a sky map, and Figure 
\ref{fig:spectindexregion} as a function of region number.

Several interesting features can be seen in Figure
\ref{fig:spectralindexwmap}. First, as already reported in the 
literature \citep[e.g.,][]{kogut:2007,dunkley:2009,macellari:2011}, 
we see that the synchrotron spectral index is steeper at high Galactic 
latitudes than along the Galactic plane. Adopting a weighted average 
over Galactic and high-latitude regions, we find mean spectral indices of
$\beta_{\textrm{plane}}=-2.98\pm0.01$ and $\beta_{\textrm{high-lat}}=-3.12\pm0.04$, 
respectively; the full-sky weighted mean is
$\beta^{\textrm{all-sky}} = -2.99\pm0.01$, being strongly dominated by
the Galactic plane regions. 

Second, we note that the spectral index along the Galactic plane
appears steeper toward the Galactic center and anticenter ($l =
0^{\circ}$ and $180^{\circ}$) than toward the Galactic spiral arms
($l=90^{\circ}$ and $-90^{\circ}$). This becomes even more clear in
Figure \ref{fig:spectindexlongitude}, in which we plot the 
algorithm-averaged synchrotron spectral index for the Galactic plane regions,
ordered according to Galactic longitude. We fit two different models
to these data points, namely a constant, $\beta_{1}(l) = c_1$, and an
offset sine function, $\beta_2(l) = c_2 + a\sin(2l - 90^{\circ})$,
using a simple $\chi^2$ minimization routine. The resulting best-fit
parameters are $c_1=-2.98$ for model 1, and $(c_2,a) = (-2.85,0.17)$
for model 2, with $\chi^2$s of 41 and 2.7 for 9 and 8 degrees of
freedom, respectively. The corresponding probabilities-to-exceed (PTE)
are $10^{-5}$ for model 1 and 0.95 for model 2; the offset sine
function is a dramatically better fit than a pure constant.

\section{\emph{WMAP} constraints on synchrotron emission in the BICEP2 field}

\begin{figure*}[t]
\begin{center}
\mbox{\epsfig{figure=fig/WMAP9_bicep_1deg.pdf,width=0.7\linewidth,clip=}}
\mbox{\epsfig{figure=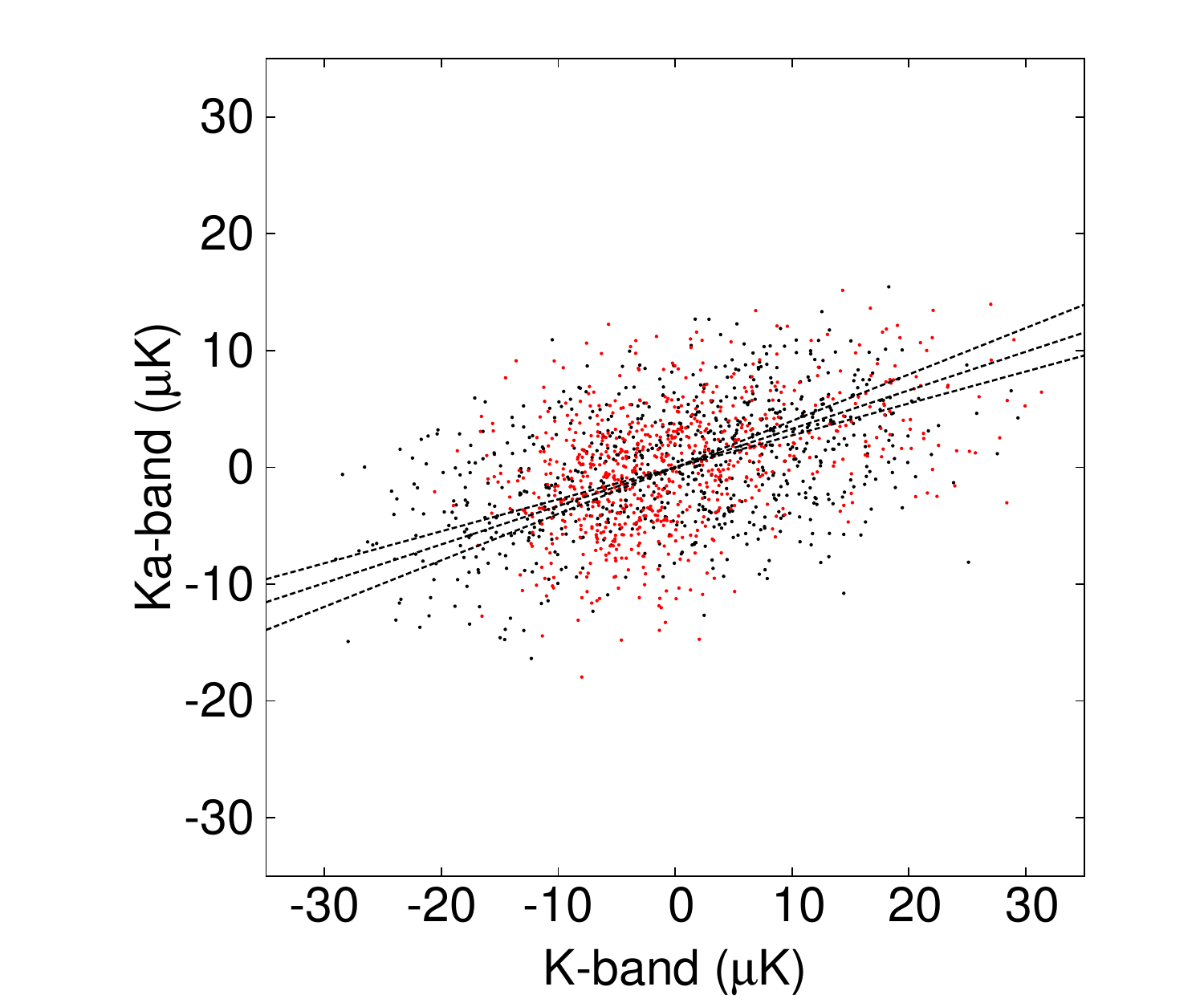,width=0.45\linewidth,clip=}}
\mbox{\epsfig{figure=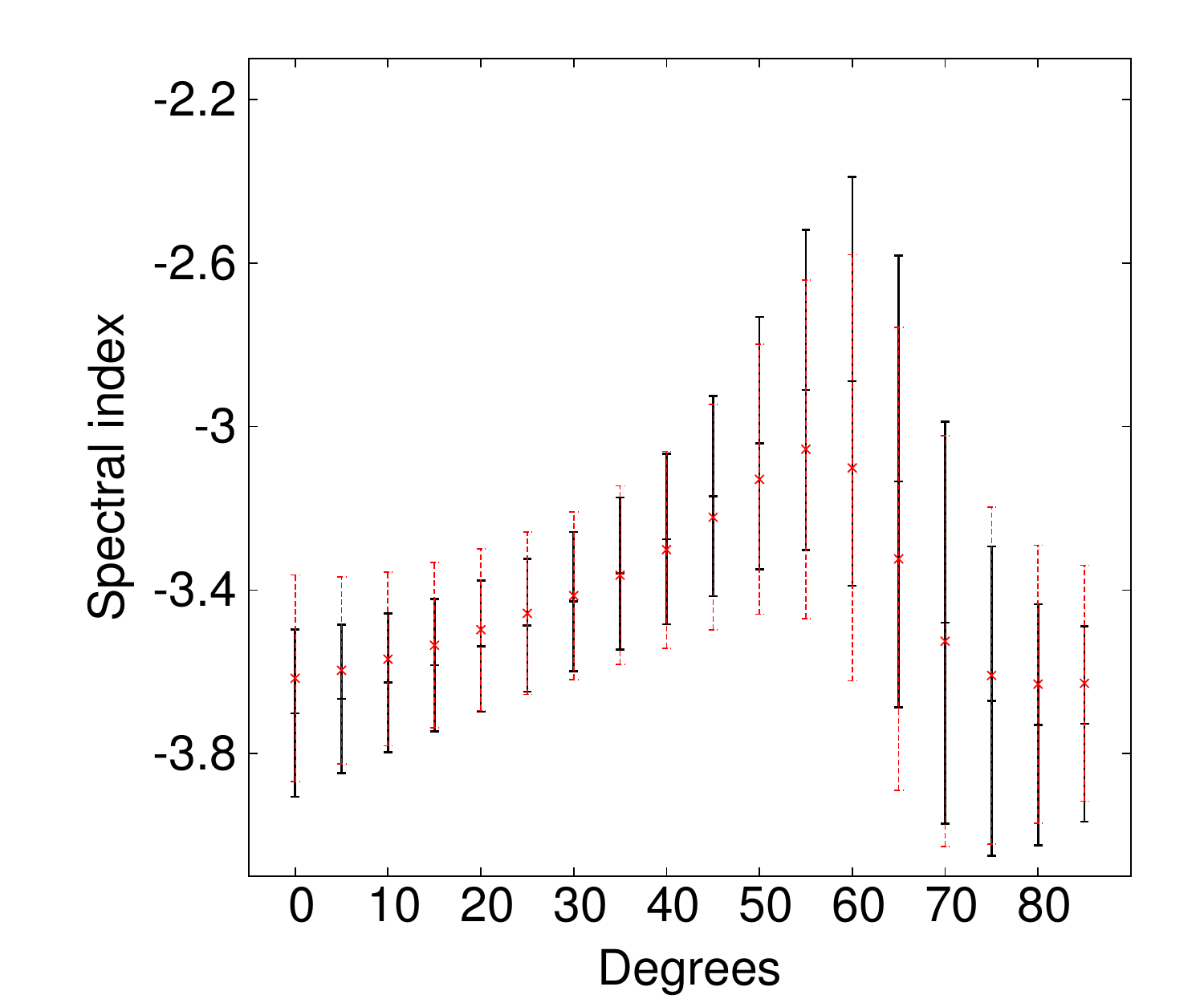,width=0.45\linewidth,clip=}}
\end{center}
\caption{Top: \emph{WMAP} K- and Ka-band polarization maps at the
  BICEP2 field, plotted in Galactic coordinates and smoothed to
  $1^{\circ}$ FWHM. Bottom left: T--T scatter plot between the
  two channels. From top to bottom, the dashed lines correspond to
  spectral indices of $\beta=-2.5$, $-3.0$, and $-3.5$,
  respectively. Bottom right: synchrotron spectral index as a
  function of polarization orientation, evaluated using the T--T plot
  (black) and ML (red, dashed) techniques.}
\label{fig:bicep}
\end{figure*}

In 2014 March, \citet{bicep2:2014} claimed the first detection of
large-scale B-mode CMB polarization, after observing an exceptionally
clean region of the southern sky for three years. The BICEP2 field is
defined roughly in terms of a rectangle given by $-40^{\circ} \lesssim
\textrm{R.A.} \lesssim 40^{\circ}$, $-65^{\circ} \lesssim \textrm{decl.}
\lesssim -50^{\circ}$ in celestial coordinates and is situated
within a larger particularly low foreground region known as the
``southern hole''. The claimed amplitude of the B-mode power excess
was larger than many had anticipated, with a tensor-to-scalar ratio of
$r=0.20^{+0.07}_{-0.05}$, corresponding to a map domain B-mode
amplitude of $0.2\,\mu\textrm{K}$. However, while this measurement
formally corresponds to a $7\,\sigma$ rejection of the null hypothesis
of no excess signal, the BICEP2 could only rule out a synchrotron-based 
explanation at the $2.3\,\sigma$ significance level using BICEP
data alone.

However, adopting a synchrotron spectral index of $\beta=-3.3$ and
extrapolating the low-$\ell$ K-band angular power spectrum to degree
scales, the BICEP2 team derived an upper limit on the residual
synchrotron contamination of $r=0.003$ at 150~GHz. Using the machinery
presented in this paper, we are in the position of understanding some
of the uncertainties associated with this projection. First, in the
top panel of Figure \ref{fig:bicep} we show the BICEP2 field of the
\emph{WMAP} K and Ka bands, smoothed to $1^{\circ}$ FWHM. Here one can
clearly see by eye large-scale synchrotron emission with an amplitude
up to $30-50\,\mu\textrm{K}$ in K band, dropping to a maximum of
$10-15\,\mu\textrm{K}$ in Ka band. Both maps are clearly noise
dominated on $1^\circ$ scales.

In the bottom left panel of Figure \ref{fig:bicep}, we show the T--T
scatter plot between the two maps. The dashed lines correspond to
spectral indices of $\beta=-2.5$, $-3.0$, and $-3.5$, respectively; with
the amount of noise present in these data, it is highly nontrivial to
determine by eye which line is the best fit, even for such a wide
range of spectral indices. This observation is made more quantitative
in the bottom right panel of the same figure, which shows the spectral
index as a function of polarization orientation, similar to those
shown in Figure \ref{fig:alphaplots}. Here we see that the allowed
spectral index range is indeed large, spanning from roughly $-3.8$ to
$-2.5$. To sum up, it seems clear that the \emph{WMAP} polarization data are
simply not sufficiently sensitive to allow a robust measurement of the
synchrotron emission in this region, neither in terms of amplitude nor
spectral index.

Instead, we need to resort to simpler extrapolations. One estimate can
be derived from the standard deviation of the K-band map. After
removing all multipoles below $\ell \leq 25$, to which BICEP2 is not
sensitive, and smoothing to $1^{\circ}$ FWHM, the observed K-band
standard deviation is $7.5\,\mu\textrm{K}$ over the BICEP2 field. The
predicted noise standard deviation from the \emph{WMAP} noise
characterization is $7.1\,\mu\textrm{K}$, computed from simulations
filtered the same way as the observations. Under the assumption that
the signal and noise are statistically independent and add in
quadrature, the predicted synchrotron standard deviation is therefore
$\sqrt(7.5^2-7.1^2) = 2.4\,\mu\textrm{K}$ over the relevant multipole
range. Scaling this to 150~GHz with a spectral index of $\beta=-3.12$
(see Section \ref{sec:results}), and accounting for the conversion
factor between antenna to thermodynamic temperature, we find that an
expected synchrotron signal at 150~GHz of
\begin{equation}
\sigma_{\textrm{max}} = 2.4\,\mu\textrm{K}\cdot
\left(\frac{150}{22.45}\right)^{-3.12} \cdot \frac{1.73}{1.01} =
0.011\,\mu\textrm{K}.
\label{eq:synchamp}
\end{equation}
For comparison, the standard deviation of a pure B-mode signal with
$r=0.2$ (0.003) is $0.08\,\mu\textrm{K}$ ($0.01\,\mu\textrm{K}$). Thus,
from our calculation it appears that the most likely synchrotron
contamination in the BICEP2 tensor-to-scalar ratio is indeed
$r=0.003$. Note, though, that our value is a predicted bias, not an
upper limit.

In the above calculation, we have assumed an average high-latitude
synchrotron spectral index of $\beta=-3.12$. However, as seen in the
bottom panel of Figure \ref{fig:bicep}, the data do allow the index to
be substantially flatter, because of the particularly high noise in
this region. In the very worst case scenario, the index could be
$\beta=-2.5$. Inserting this index into Equation \eqref{eq:synchamp}
yields a synchrotron rms value of $0.036\,\mu\textrm{K}$. Again adding
signals in quadrature, we find that synchrotron contamination can in
the absolute worst case scenario make up at most 20\% of the signal
detected by BICEP2; and if the synchrotron properties in the BICEP2
field are anything similar to the rest of the sky, except for
amplitude, we expect it to be on the order of 2\%.
We conclude that ``vanilla''synchrotron contamination is not a 
promising candidate to explain the BICEP2 power excess.

\section{Conclusions}
\label{sec:conclusions}

We have measured the spectral index of polarized synchrotron emission
from the 9 yr \emph{WMAP} K and Ka bands. We have implemented two different
methods, one traditional T--T plot method and one ML
based method. We partitioned the sky into 24 disjoint regions,
excluding particularly bright point sources and the Galactic center,
and estimated a spectral index for each region. For the full sky, we
find an overall inverse-variance weighted spectral index of
$\beta^{\textrm{all-sky}} = -2.99\pm0.01$. Considering the Galactic
plane and high-latitude regions separately, the two weighted means are
$\beta^{\textrm{plane}} = -2.98\pm0.01$ and $\beta^{\textrm{high-lat}}
= -3.12\pm0.04$. Thus, we find that the spectral index flattens by
$0.14$ from the Galactic plane to high latitudes, in good
agreement with previous analyses \citep[e.g.,][]{kogut:2007}.

Considering only the Galactic plane regions, we additionally observe a
noticeable trend of steeper spectral indices toward the Galactic
center and anticenter than toward the Galactic spiral arms. Fitting
an offset sinusoidal to the data, we find a best-fit model of the form
$\beta(l) = -2.85 + 0.17\sin(2l - 90^{\circ})$.
Overall, there seems to be substantial evidence for
spatial variation of the synchrotron spectral index.

Finally, we comment on the possibility of explaining the recent BICEP2
measurements of B-mode polarization in terms of synchrotron
contamination. Overall, we reach similar conclusions to those
presented by BICEP2, albeit with slightly more conservative numbers:
We find that the most likely bias from synchrotron contamination
in the BICEP2 field corresponds to a tensor-to-scalar
ratio of $r=0.003$. In the absolute worst case scenario, when
assuming a synchrotron spectral index of $\beta=-2.5$, which is the
flattest index allowed by the data in this region, and significantly
flatter than the rest of the sky, at most 20\% of the observed signal
can be explained in terms of synchrotron emission. However, before
dismissing synchrotron completely, it is worth making one caveat:
these calculations assume that synchrotron emission follows a perfect
power law from 23 to 150~GHz. If there is a significant positive
curvature in the synchrotron spectrum, these conclusions clearly 
would have to be revised. 

\begin{acknowledgements}
The computations presented in this paper were carried out on Abel, a
cluster owned and maintained by the University of Oslo and NOTUR.
This project was supported by the ERC Starting Grant StG2010-257080.
I.K.W. acknowledges support from ERC grant 259505.  Part of the research
was carried out at the Jet Propulsion Laboratory, California Institute
of Technology, under a contract with NASA.  Some of the results in
this paper have been derived using the HEALPix \citep{gorski:2005}
software and analysis package.
\end{acknowledgements}

\end{document}